\documentclass[aps,prd,10pt,         
               preprintnumbers,numbers,sort&compress,nofootinbib,
                            showpacs,
               colorlinks,
               linkcolor=blue,
               citecolor=blue]{revtex4-1}
   \newcommand{\exclude}[1]{}

\usepackage{graphicx,amsmath,amssymb,bm}
\usepackage{psfrag}
\usepackage{feynmp}
\usepackage{hyperref}
\usepackage{enumitem}
\newcommand{\nn}{\nonumber}
\newcommand{\beq}{\begin{equation}}
\newcommand{\eeq}{\end{equation}}
\newcommand{\be}{\begin{eqnarray}}
\newcommand{\ee}{\end{eqnarray}}

\def\dd{ \,\mathrm{d} }

\def\+{\dagger}
\def\la{\langle}
\def\ra{\rangle}
\def\<{\langle}
\def\>{\rangle}

\newcommand{\Lqcd}{\Lambda_{\mathrm{QCD}}}

\begin{document}

\title{ Topological Susceptibility  and Contact Term in QCD.   A Toy Model.}

\author{ Evan Thomas \& Ariel R. Zhitnitsky} 
 \affiliation{Department of Physics \& Astronomy, University of British Columbia, Vancouver, B.C. V6T 1Z1, Canada}

\begin{abstract}
We study a number of different ingredients related to $\theta$ dependence, the non-dispersive contribution in topological susceptibility with the ``wrong'' sign, topological sectors in gauge theories, and related subjects using a simple ``deformed QCD''. This model is a weakly coupled gauge theory, which however has all the relevant essential elements allowing us to study difficult and nontrivial questions which are known to be present in real strongly coupled QCD. Essentially we want to test the ideas related to the $U(1)_A$ problem in a theoretically controllable manner using the ``deformed QCD" as a toy model. One can explicitly see microscopically how the crucial elements work.
\end{abstract}

\maketitle

\section{Introduction and motivation} \label{introduction}

In the present paper, we study the topological susceptibility in a ``deformed'' QCD, wherein we can work analytically.  The goal is to understand   a number of different ingredients related to $\theta$ dependence, the non-dispersive contribution in topological susceptibility with the ``wrong sign", topological sectors in gauge theories, and many related subjects using a simple ``deformed QCD". This model is a weakly coupled gauge theory, which however has all essential elements allowing us to study difficult and nontrivial questions which are known to be present in real strongly coupled QCD. Essentially we want to test the ideas related to the $U(1)_A$ problem formulated long ago \cite{witten,ven,vendiv,Rosenzweig:1979ay,Nath:1979ik,Kawarabayashi:1980dp}  in a theoretically controllable manner using the ``deformed QCD" \cite{Yaffe:2008} as a toy model. One can explicitly see microscopically how all the crucial elements work.
For early studies on relation between topology, long range order,  and presence of unphysical non-propagating degrees of freedom see \cite{Luscher:1978rn}. 

As a result of these studies we find  that  the QCD vacuum energy density contains a contribution, that
is not related to the physical   propagating degrees of freedom in the system, but rather originates from nontrivial topological sectors of the theory. 
Our findings  shed some light on microscopic mechanisms related to the resolution of the old $U(1)_A$ problem \cite{witten,ven,vendiv,Rosenzweig:1979ay,Nath:1979ik,Kawarabayashi:1980dp}. While numerically the Monte Carlo simulations  support the picture developed in \cite{witten,ven,vendiv,Rosenzweig:1979ay,Nath:1979ik,Kawarabayashi:1980dp}, there were no analytical computations which would provide hints as to the microscopic nature of the contact term with the ``wrong sign". Still, 30 years after the resolution of the $U(1)_A$  problem, the nature and the source of the contact term remains a mystery. The goal of this paper is to fill this gap, to identify the gauge configurations which saturate the contact term.

The strategy for the present study is as follows. We start with analysis of the non-dispersive ``contact'' term  in a completely solvable $2$-dimensional gauge theory wherein there are not any physical propagating degrees of freedom. We explain how the contact terms emerges in this case. We then consider a simplified (``deformed'') version of QCD which, on one hand, is a weakly coupled gauge theory wherein computations can be performed in theoretically controllable manner. On other hand, this deformation preserves all relevant elements such as confinement, degeneracy of topological sectors, nontrivial $\theta$ dependence, presence of non-dispersive contribution to topological susceptibility, and other crucial aspects, for this phenomenon to emerge. This ``deformed'' theory has recently been developed \cite{Yaffe:2008} and will be further discussed in section \ref{deformedqcd}. Finally, we compute the contact term in four dimensional ``deformed QCD"  in section \ref{section-chi} where we interpret the contact term in terms of tunnelling events between different topological sectors of the gauge theory. We explain how the Ward Identities are automatically satisfied when light quark is inserted into the system. 
     
\section{The contact term and degeneracy of topological sectors}\label{2d}

In this section we present an overview of the nature of the contact term which is not related to any physical degrees of freedom. We explain how we know about its mere existence and how the anomalous Ward Identities (WI) require its presence for the gauge theory to be self-consistent. We also give a simple two dimensional example explaining how this term emerges in gauge theories. Put differently, the nature of this ``weird" contribution is entirely determined by the topological properties of the model rather than the physical propagating degrees of freedom of the system. Thus, we will find similar behaviour in theories with similar topological properties irrespective of the particularities of the theories.

\subsection{The contact term}\label{contacttermsection}

We start with  definition of the topological susceptibility $\chi$ which is the main ingredient of the resolution of the $U(1)_A$ problem in QCD \cite{witten,ven,vendiv}, see also~\cite{Rosenzweig:1979ay,Nath:1979ik,Kawarabayashi:1980dp}. The necessity for the contact term in topological susceptibility $\chi$ can be explained in few lines as follows. We define the topological susceptibility $\chi$ in the standard way:
\be
\label{chi}
 \chi (\theta=0)=  \left. \frac{\partial^2E_{\mathrm{vac}}(\theta)}{\partial \theta^2} \right|_{\theta=0}=  \lim_{k\rightarrow 0} \int \!\dd^4x e^{ikx} \la T\{q(x), q(0)\}\ra ,
 \ee
where $\theta$ is the conventional $\theta$ parameter which enters the Lagrangian along with topological density operator $q (x)$, see precise definitions below. The most important feature of the topological susceptibility $\chi$, for our present discussion, is that it does not vanish in spite of the fact that $q=\partial_{\mu}K^{\mu}$ is total divergence. Furthermore, any physical state of mass $m_G$, momentum $k\rightarrow 0$  and coupling $\la 0| q| G\ra= c_G$ contributes to the dispersive portion of the topological susceptibility with negative sign\footnote{We use the Euclidean notations  where  path integral computations are normally performed.}
\be	\label{G}
  \chi_{\rm dispersive} \sim  \lim_{k\rightarrow 0} \int \!\dd^4x e^{ikx} \la T\{q(x), q(0)\}\ra  \sim  
    ~\lim_{k\rightarrow 0}   \frac{\la  0 |q|G\ra \la G| q| 0\ra }{-k^2-m_G^2}\simeq -\frac{|c_G|^2}{m_G^2} \leq 0, 
\ee
while the resolution of the $U(1)_A$ problem, which would provide a physical mass for the $\eta'$ meson, requires a positive sign for the topological susceptibility (\ref{top1}), see the original reference~\cite{vendiv} for a thorough discussion, 
\be	\label{top1}
  \chi_{\rm non-dispersive}= \lim_{k\rightarrow 0} \int \!\dd^4x e^{ikx} \la T\{q(x), q(0)\}\ra > 0 \, .
\ee
Therefore, there must be a contact contribution to $\chi$, which is not related to any propagating physical degrees of freedom, and it must have the ``wrong'' sign, by which we mean opposite to any term originating from physical propagators, in order to saturate the topological susceptibility (\ref{top1}). In the framework \cite{witten} the contact term with the ``wrong'' sign has been simply postulated, while in refs.\cite{ven,vendiv} the Veneziano ghost had been introduced to saturate the required property (\ref{top1}). This Veneziano ghost field is simply an unphysical degree of freedom with the ``wrong'' sign in the kinetic term such that it generates the same contact term when integrated out. It should be emphasized that these two descriptions are equivalent and simply two separate ways of describing the same physics and that in these two pictures, the claim that the ``contact'' term does not come from physical propagating degrees of freedom is manifest. 

It should be mentioned here that the ``wrong'' sign in topological susceptibility (\ref{top1}) is not the only manifestation of this ``weird'' unphysical  degree of freedom. In fact, one can argue that the well known mismatch between  Bekenstein-Hawking entropy and the entropy of entanglement for gauge fields is due to the same gauge configurations which saturate the contact term in the topological susceptibility in QCD~\cite{Zhitnitsky:2011tr}. In both cases the extra term with a ``wrong'' sign is due to distinct topological sectors in gauge theories. This extra term has non-dispersive nature, can not be restored from the conventional spectral function through dispersion relations, and can not be associated with any physical propagating degrees of freedom.

\subsection{Topological susceptibility and contact term in 2d QED}\label{contact} 
  
The goal here is to give some insights on the nature of the contact term using a simple exactly solvable two dimensional $QED_2$ ~\cite{KS}. We follow \cite{Zhitnitsky:2011tr,Zhitnitsky:2010ji}  to discuss all essential elements  related to the contact term.

We start by considering two dimensional photodynamics (without fermions) which is naively a trivial theory as it does not have any physical propagating degrees of freedom. However, we shall argue that this (naively trivial) two dimensional photodynamics nevertheless has a contact term which is related to the existence of  different topological sectors in the theory. Thus, the presence of degenerate topological sectors in the system, which we call the ``degeneracy" for short\footnote{Not to be confused with conventional term ``degeneracy" when two or more physically distinct states are present in the system. In the context of this paper, the ``degeneracy" references the existence of winding states $| n\ra$ constructed as follows: ${\cal T} | n\ra= | n+1\ra$.  In this formula the operator ${\cal T}$ is  the  large gauge transformation operator  which commutes  with the Hamiltonian $[{\cal T}, H]=0$, implying the ``degeneracy" of the winding states $| n\ra$. The physical vacuum state is {\it unique} and constructed as a superposition of $| n\ra$ states. In path integral approach the presence of $N$ different sectors in the system is reflected  by  summation over $ { k \in \mathbb{Z}}$ in eqs. (\ref{Z}, \ref{chi3}, \ref{chi4}). }, is the source for this contact term which is not related to any physical propagating degrees of freedom.

The topological susceptibility $\chi$ in this model is defined as follows 
\be	\label{chi1}
  \chi \equiv \frac{e^2}{4\pi^2} \lim_{k\rightarrow 0} \int \dd^2x e^{ikx}\left< T E(x) E(0) \right> ,
\ee
where $q=\frac{e}{2\pi}E$ is the topological charge density operator and 
\be	\label{k}
  \int  \dd^2x ~q(x)= \frac{e}{2\pi} \int \dd^2x ~E(x) =k
\ee
is the integer valued topological charge in the 2d $U(1)$ gauge theory, $E(x)=\partial_1A_2-\partial_2A_1$ is the field strength. The expression for the topological susceptibility in 2d Schwinger QED model when the fermions are included into the system is known exactly \cite{SW} 
\be	\label{exact}
  \chi_{QED}= \frac{e^2}{4\pi^2}  \int   \dd^2x \left[ \delta^2(x) - \frac{e^2}{2\pi^2} K_0(\mu |x|) \right] ,
\ee
where $\mu^2=e^2/\pi$ is the mass of the single physical state in this model, and $K_0(\mu |x|) $ is the modified Bessel function of order $0$, which is the Green's function of this massive particle. The expression for $\chi$ for pure photodynamics is given by (\ref{exact}) with coupling  $e=0$ in the brackets\footnote{the factor $\frac{e^2}{4\pi^2}$ in front  of (\ref{exact}) does not vanish in this limit as it is due to our definition (\ref{chi1}) rather than result of dynamics} which corresponds to the de-coupling from matter field $\psi$, i.e.
\be	\label{exact1}
  \chi_{E\&M}= \frac{e^2}{4\pi^2}  \int   \dd^2x \left[ \delta^2(x)  \right]= \frac{e^2}{4\pi^2}.
\ee
The crucial observation here is as follows: any physical state contributes to $\chi$ with negative sign 
\be	\label{dispersion}
  \chi_{dispersive} \sim  \lim_{k\rightarrow 0} \sum_n  \frac{\la 0| \frac{e}{2\pi}   E  |n\ra \la n | \frac{e}{2\pi}   E |0\ra }{-k^2-m_n^2} <0,
\ee
in accordance with the general formula (\ref{G}) in four dimensions discussed previously. In particular, the term proportional $ -K_0(\mu |x|) $ with negative sign in eq. (\ref{exact}) results from the only physical field of mass $\mu$. However, there is also a contact term $ \int   \dd^2x \left[ \delta^2(x)  \right]$ in (\ref{exact}) and (\ref{exact1}) which contributes to the topological susceptibility $\chi$  with the {\it opposite sign}, and which can not be identified according to (\ref{dispersion}) with any contribution from any physical asymptotic state. In the two-dimensional theory without a fermion (photodynamics), there are no asymptotic states since there are no possible polarization states, and so it is clear that the contact term (\ref{exact1}) is not related to any physical propagating degree of freedom. Likewise, with a fermion included, there is one physical degree of freedom, yet we see also the additional ``contact'' contribution in (\ref{exact}).

This term has  fundamentally different, non-dispersive  nature. In fact it is ultimately related to different topological sectors of the theory and the degeneracy of the ground state as we shortly review below. Without this contribution it would be impossible to satisfy the Ward Identity (WI) because the physical propagating degrees of freedom can only contribute with sign $(-)$ to the correlation function as (\ref{dispersion}) suggests, while the WI requires $\chi_{QED}(m=0)= 0$  in the  chiral limit $m=0$. One can explicitly check that WI is indeed automatically satisfied  only as a result of exact cancellation between conventional dispersive term with sign $(-)$ and non-dispersive term (\ref{exact1}) with sign $(+)$,   
\be	\label{chi2}
  \chi_{QED}  = \frac{e^2}{4\pi^2}  \int   \dd^2x \left[ \delta^2(x) - \frac{e^2}{2\pi^2} K_0(\mu |x|) \right]  
    = \frac{e^2}{4\pi^2} \left[ 1- \frac{e^2}{\pi}\frac{1}{\mu^2}\right]= \frac{e^2}{4\pi^2} \left[ 1-1\right]=0.
\ee
Therefore, contact term actually plays a crucial role in maintaining the consistency of the theory, because the WI can not be satisfied without it. While the exact formula (\ref{exact}) is known, it does not hint at the kind of physics responsible for the contact term with the ``wrong sign", mainly what sort of field configurations should saturate the contact term. Below, we provide some insights on this matter.
 
\subsection{The origin of the contact term -- summation over topological sectors}\label{sectors}

The goal here is to demonstrate that the contact term in the exact formulae (\ref{exact}) and (\ref{exact1})   is a result of the summation over different topological $k$ sectors  in the $2d$ pure $U(1)$ gauge theory. The relevant ``instanton-like" configurations are defined on a two dimensional Euclidean torus with total area $V$ as follows~\cite{SW},
\be	\label{instanton}
  A_{\mu}^{(k)}=-\frac{\pi k}{e V}\epsilon_{\mu\nu}x^{\nu}, ~~~ e E^{(k)}=\frac{2\pi k}{V}, 
\ee
such that the action of this classical configuration is
\be	\label{action}
  \frac{1}{2}\int d^2x E^2= \frac{2\pi^2 k^2}{e^2 V}.
\ee
This configuration corresponds to the topological charge $k$ as defined by (\ref{k}). The next step is to compute the topological susceptibility for the theory defined by the following partition function
\be	\label{Z}
  {\cal{Z}}=\sum_{ k \in \mathbb{Z}}{\int {\cal{D}}}A {e^{-\frac{1}{2}\int d^2x E^2}}.
\ee
All integrals in this partition function are gaussian and can be easily evaluated.
The result is determined  essentially  by the classical configurations (\ref{instanton}) and (\ref{action}) since real propagating degrees of freedom are not present in the system of pure $U(1)$ gauge field theory in two dimensions. We are interested in computing $\chi$ defined by eq. (\ref{chi1}). In the path integral approach it can be represented as follows, 
\be	\label{chi3}
  \chi_{E\&M}=\frac{e^2}{4\pi^2 \cal{Z}}\sum_{k\in \mathbb{Z}}{\int{\cal{D}}}A\int\dd^2x E(x) E(0){e^{-\frac{1}{2}\int d^2x E^2}}.  
\ee
This gaussian integral can be easily evaluated  and  the result is as follows~\cite{Zhitnitsky:2010ji, Zhitnitsky:2011tr}, 
\be	\label{chi4}
  \chi_{E\&M}= \frac{e^2}{4\pi^2} \cdot V\cdot \frac{ \displaystyle \sum_{ k \in \mathbb{Z}} \frac{4\pi^2k^2}{e^2 V^2}  \exp(-\frac{2\pi^2 k^2}{e^2 V})}{ \displaystyle \sum_{ k \in \mathbb{Z}}  \exp(-\frac{2\pi^2 k^2}{e^2 V})}.
\ee
In the large volume limit $V\rightarrow \infty$ one can evaluate the sums entering (\ref{chi4}) by replacing $ \sum_{ k \in \mathbb{Z}}\rightarrow \int d k $ such that the leading term in eq. (\ref{chi4}) takes the form
\be	\label{chi5}
  \chi_{E\&M} = \frac{e^2}{4\pi^2} \cdot V\cdot  \frac{4\pi^2}{e^2 V^2} \cdot\frac{e^2 V}{4\pi^2}=  \frac{e^2}{4\pi^2}.
\ee
A few comments are in order. First, the obtained expression for the topological susceptibility (\ref{chi5}) is finite in the  limit $V\rightarrow \infty$, coincides with the contact term from exact computations (\ref{exact}), (\ref{exact1}) performed for the $2d$ Schwinger model, and has the ``wrong" sign in comparison with any physical contributions (\ref{dispersion}). Second, the topological sectors with very large $k\sim \sqrt{e^2V}$ saturate the series (\ref{chi4}). As one can see from the computations presented above, the final result  (\ref{chi5}) is sensitive to the boundaries, infrared regularization, and many other aspects which are normally ignored when a  theory from the very beginning is formulated in infinite space with  conventional  assumption about  trivial behaviour at infinity. Last, but not least: the contribution (\ref{chi5}) does not vanish in a trivial model with no propagating degrees of freedom present in the system. This term is entirely determined by the behaviour at the boundary, which is conveniently represented by the classical topological configurations (\ref{instanton}) describing  different topological sectors (\ref{k}), and accounts for the degeneracy of the ground state. In this way, large distance physics enters despite the lack of physical long distance degrees of freedom. Furthermore, ee know that this term must be present in the theory when the dynamical quarks are introduced into the system. Indeed, it plays a crucial role in this case as it saturates the WI as (\ref{chi2}) shows.

We conclude this section by noting that the contact term in the framework~\cite{KS} can be computed in terms of the Kogut-Susskind ghost by replacing the standard path integral procedure of summation over different topological sectors above as follows. The topological density $q=\frac{e}{2\pi}E$ in $2d$ QED is given by $\frac{e}{2\pi} E= (\frac{e}{2\pi}){\frac{\sqrt{\pi}}{e}}\left(  \Box\hat\phi -  \Box\phi_1 \right)$ where $\hat\phi$ is the physical massive field of the model and $\phi_1$ is the ghost~\cite{KS}. The relevant correlation function in coordinate space which enters the expression for the topological susceptibility (\ref{chi1}) can be explicitly computed using the ghost as follows 
\be	\label{chi6}
  \chi_{QED}(x) &\equiv& \left< T   \frac{e}{2\pi} E(x) , \frac{e}{2\pi} E(0) \right>    
    =\left( \frac{e}{2\pi} \right)^2\frac{\pi}{e^2} \int \frac{\dd^2p}{\left(2\pi\right)^2} 
    p^4 e^{-i p x} \left[ - \frac{1}{p^2+\mu^2} + \frac{1}{p^2} \right] \nn \\
  & = & \left(\frac{e}{2\pi} \right)^2\left[ \delta^2(x) - \frac{e^2}{2\pi^2} K_0(\mu |x|)\right]    
\ee
where we used the known expressions for the  Green's functions. The obtained expression precisely reproduces the exact result (\ref{exact}) as claimed.  In the limit $e\rightarrow 0$ when  the fermion matter  field decouples from gauge degrees of freedom  we reproduce the contact term  (\ref{exact1}, \ref{chi5}) which was previously derived as a result of summation over different topological sectors of the theory. The non-dispersive contribution manifests itself in this description in terms of the unphysical ghost scalar field which provides the required ``wrong" sign for the contact term. These two different descriptions are analogous to the same two computations in four dimensions mentioned in section \ref{contacttermsection}, with and without the Veneziano ghost, and again we emphasize the equivalence of the two. In the picture wherein the contact term is saturated by a ghost field we see again how the contact term is related not to physical propagating degrees of freedom.

\section{Deformed QCD} \label{deformedqcd}

Next we discuss the ``center-stablized" deformed Yang-Mills developed in \cite{Yaffe:2008} and references therein, before moving on to a discussion of the topological properties of this theory in section \ref{section-chi}. In the deformed theory an extra term is put into the Lagrangian in order to prevent the center symmetry breaking that characterizes the QCD phase transition between ``confined" hadronic matter and ``deconfined" quark-gluon plasma. Thus we have a theory which remains confined at high temperature in a weak coupling regime, and for which it is claimed \cite{Yaffe:2008} that there does not exist an order parameter to differentiate the low temperature (non-abelian) confined regime from the high temperature (abelian) confined regime. We follow \cite{Yaffe:2008} in deriving the relevant parts of the theory.

\subsection{Formulation of the theory}

We start with pure Yang-Mills (gluodynamics) with gauge group $SU(N)$ on the manifold $\mathbb{R}^{3} \times S^{1}$ with the standard action
\be \label{standardYM}
	S^{YM} = \int_{\mathbb{R}^{3} \times S^{1}} d^{4}x\; \frac{1}{2 g^2} \mathrm{tr} \left[ F_{\mu\nu}^{2} (x) \right],
\ee
and add to it a deformation action,
\be \label{deformation}
	\Delta S \equiv \int_{\mathbb{R}^{3}}d^{3}x \; \frac{1}{L^{3}} P \left[ \Omega(\mathbf{x}) \right],
\ee 
built out of the Wilson loop (Polyakov loop) wrapping the compact dimension,
\be \label{loop}
	\Omega(\mathbf{x}) \equiv \mathcal{P} \left[ e^{i \oint dx_{4} \; A_{4} (\mathbf{x},x_{4})} \right].
\ee
The "double-trace" deformation potential $P \left[ \Omega \right]$ respects the symmetries of the original theory and is built to stabilize the phase with unbroken center symmetry. It is defined by
\be \label{deformationpotential}
	P \left[ \Omega \right] \equiv \sum_{n = 1}^{\lfloor N/2 \rfloor} a_{n} \left| \mathrm{tr} \left[ \Omega^{n} \right] \right| ^{2}.
\ee
Here $\lfloor N/2 \rfloor$ denotes the integer part of $N/2$ and $\left\{ a_{n} \right\}$ is a set of suitably large positive coefficients.

The first term of $P \left[ \Omega \right]$,  proportional to $\left| \mathrm{tr} \left[ \Omega \right] \right|^{2}$ (with a sufficiently large positive coefficient), will prevent breaking of the center symmetry from $\mathbb{{Z}}_{N}$ to $\mathbb{{Z}}_{1}$ with order parameter $\langle \mathrm{tr} \left[ \Omega \right] \rangle$, but will not prevent $\mathrm{tr} \left[ \Omega^{2} \right]$ from developing a vacuum expectation value so that it will not prevent the center symmetry breaking from $\mathbb{{Z}}_{N}$ to $\mathbb{{Z}}_{2}$ with order parameter $\langle \mathrm{tr} \left[ \Omega^{2} \right] \rangle$. The term proportional to $\left| \mathrm{tr} \left[ \Omega^{2} \right] \right|^{2}$ however does prevent such a symmetry braking. Likewise, for each other subset $\mathbb{{Z}}_{p}$ of $\mathbb{{Z}}_{N}$ (with $N \; \mathrm{mod} \; p = 0$), there needs to be a corresponding term, proportional to $\left| \mathrm{tr} \left[ \Omega^{p} \right] \right|^{2}$, in the deformation potential. This is the reason for including terms up to $\left| \mathrm{tr} \left[ \Omega^{\lfloor N/2 \rfloor} \right] \right|^{2}$. Note that for real life QCD the gauge group is $SU(3)$ and so only one term is necessary,
\be
	P \left[ \Omega \right] = a \left| \mathrm{tr} \left[ \Omega \right] \right| ^{2}.
\ee

In undeformed pure gluodynamics the effective potential for the Wilson loop is minimized for $\Omega$ an element of $\mathbb{{Z}}_{N}$. The deformation potential (\ref{deformationpotential}) with sufficiently large $\{ a_{n} \}$ however changes the effective potential for the Wilson line so that it is minimized instead by configurations in which $\mathrm{tr} \left[ \Omega^{n} \right] = 0$, which in turn implies that the eigenvalues of $\Omega$ are uniformly distributed around the unit circle. Thus, the set of eigenvalues is invariant under the $\mathbb{{Z}}_{N}$ transformations, which multiply each eigenvalue by $e^{2 \pi i k / N}$ (rotate the unit circle by $k/N$). The center symmetry is then unbroken by construction. The coefficients, $\{ a_{n} \}$, can be suitably chosen such that the deformation potential, $P\left[ \Omega \right]$, forces unbroken symmetry at any compactification scales \cite{Yaffe:2008}, but for our purposes we are only interested in small compactifications ($L \ll \Lambda^{-1}$ where $L$ is the length of the compactified dimension and $\Lambda$ is the QCD scale). At small compactification, the gauge coupling at the compactification scale is small so that we can work in a perturbative regime and explicitly evaluate the potential for the Wilson loop due to Quantum fluctuations as in \cite{Gross:1981, Yaffe:2008}. The one-loop potential is
\be \label{Wilsonpotential}
	V \left[ \Omega \right] = \int_{\mathbb{R}^{3}} \; d^{3}x \frac{1}{L^{4}} \mathcal{V} \left[ \Omega (x) \right],
\ee
with
\be \label{Wilsonpotentialdensity}
	\mathcal{V} \left[ \Omega \right] = -\frac{2}{\pi^{2}} \sum_{n = 1}^{\infty} \frac{1}{n^4} \left| \mathrm{tr} \left[ \Omega^{n} \right] \right|^{2}.
\ee
The undeformed potential for the Wilson loop (\ref{Wilsonpotentialdensity}) is minimized when $\Omega$ is an element of the center, $\mathbb{{Z}}_{N}$, so that $\Omega = e^{2 \pi i k / N}$. The deformation potential (\ref{deformationpotential}) must therefore overcome this one-loop potential and force $\Omega$ to not choose one particular element of $\mathbb{{Z}}_{N}$. We must choose the coefficients $\{ a_{n} \}$ to be larger than $2/\left(\pi^{2}n^{4}\right)$. A simple choice is $a_{n} = 4/\left(\pi^{2}n^{4}\right)$. With this choice, the full one-loop effective potential for the Wilson loop is minimized for $\mathrm{tr} \left[ \Omega^{n} \right] = 0$ for all $n \neq 0 \; \mathrm{mod} \; N$, indicating unbroken center symmetry.

\subsection{Infrared description}

As mentioned in the previous section, we are interested in the regime in which the compactification size is small, $L \ll \Lambda^{-1}$, and so the gauge coupling is small a the compactification scale, $g^{2}\left( 1/L \right) \ll 1$. So, in our deformed theory, the combined effective potential for the Wilson loop is the sum of (\ref{Wilsonpotentialdensity}) and (\ref{deformationpotential}), which is minimized by field configurations with
\be \label{minimumloop}
	\Omega = \mathrm{Diag} \left( 1, e^{2\pi i/N}, e^{4\pi i/N}, \dots , e^{2\pi i (N-1)/N} \right),
\ee
up to conjugation by an arbitrary element of $SU(N)$. The configuration (\ref{minimumloop}) can be thought of as braking the gauge symmetry down to its maximal Abelian subgroup, $SU(N) \rightarrow U(1)^{N-1}$. In the gauge in which $\Omega$ is diagonal, the modes of the diagonal components of the gauge field with zero momentum along the compactified dimension describe the $U(1)^{N-1}$ photons. Modes of the diagonal gauge field with non-zero momentum in the compactified dimension form a Kaluza-Klein tower and receive masses that are integer multiples of $2\pi/L$. The remaining off-diagonal components of the gauge field form a Kaluza-Klien tower of charged $W$-bosons which receive masses that are integer multiples of $2\pi/NL$. Then the lightest $W$-boson mass, $m_{W} \equiv 2\pi/NL$, describes the scale below which the dynamics are effectively Abelian.

As described in \cite{Yaffe:2008}, the proper infrared description of the theory is a dilute gas of $N$ types of monopoles, characterized by their magnetic charges, which are proportional to the simple roots and affine root of the Lie algebra for the gauge group $U(1)^{N}$. Although the symmetry breaking is $SU(N) \rightarrow U(1)^{N-1}$, it is simpler to work with $U(1)^{N}$ and, as we will see, the extra degree of freedom will completely decouple from the dynamics. The extended root system is given by the simple roots,
\begin{equation*}
\begin{array}{lclcl}
	\alpha_{1} &=& \left( 1, -1, 0, \dots, 0 \right) &=& \hat{e}_{1} - \hat{e}_{2},\\
	\alpha_{2} &=& \left( 0, 1, -1, \dots, 0 \right) &=& \hat{e}_{2} - \hat{e}_{3},\\
		     &\vdots&	&  &  \\
	\alpha_{N-1} &=& \left( 0, \dots, 0, 1, -1 \right) &=& \hat{e}_{N-1} - \hat{e}_{N},
\end{array}
\end{equation*}
and the affine root,
\begin{equation*}
\begin{array}{lclcl}

	\alpha_{N} &=& \left( -1, 0, \dots, 0 ,1 \right) &=& \hat{e}_{N} - \hat{e}_{1}. \\
\end{array}
\end{equation*}
We denote this root system by $\Delta_{\mathrm{aff}}$ and note that the roots obey the inner product relation
\be \label{dotproduct}
	\alpha_{a} \cdot \alpha_{b} = 2\delta_{a, b} - \delta_{a, b+1} - \delta_{a, b-1}.
\ee

For a fundamental monopole with magnetic charge $\alpha_{a} \in \Delta_{\mathrm{aff}}$, the topological charge is given by
\be \label{topologicalcharge}
	Q = \int_{\mathbb{R}^{3} \times S^{1}} d^{4}x \; \frac{1}{16 \pi^{2}} \mathrm{tr} \left[ F_{\mu\nu} \tilde{F}^{\mu\nu} \right]
		= \pm\frac{1}{N},
\ee
and the Yang-Mills action is given by
\be \label{YMaction}
	S_{YM} = \int_{\mathbb{R}^{3} \times S^{1}} d^{4}x \; \frac{1}{2 g^{2}} \mathrm{tr} \left[ F_{\mu\nu}^{2} \right]
		= \left| \int_{\mathbb{R}^{3} \times S^{1}} d^{4}x \; \frac{1}{2 g^{2}} \mathrm{tr} \left[ F_{\mu\nu} \tilde{F}^{\mu\nu} \right] \right|
		= \frac{8 \pi^{2}}{g^{2}} \left| Q \right|.
\ee
The second equivalence hold because the classical monopole solutions are self dual, \cite{Gross:1981}
\begin{equation*}
	F_{\mu\nu} = \tilde{F}_{\mu\nu}.
\end{equation*}
For an antimonopole with magnetic charge $-\alpha_{a}$, the Yang-Mills action is the same (\ref{YMaction}) and the topological charge changes sign, $Q = -1/N$.

So the infrared description, at distances larger than the compactification length $L$, is given by a three dimensional dilute monopole gas with $N$ types of monopoles (and so $N$ types of anti-monopoles) interacting by a species dependent Coulomb potential with interactions defined by the inner product (\ref{dotproduct}),
\be \label{Coulombpotential}
	V_{a, b} (\mathbf{r}) = L \left( \frac{2 \pi}{g} \right)^{2} \frac{(\pm \alpha_{a}) \cdot (\pm \alpha_{b})}{4 \pi \left| \mathbf{r} \right|}
		= \pm L \left( \frac{2 \pi}{g} \right)^{2} \frac{2 \delta_{a, b} - \delta_{a, b-1} - \delta_{a, b+1}}{4 \pi \left| \mathbf{r} \right|},
\ee
where the overall sign is plus for a monopole-monopole or antimonopole-antimonopole interaction and minus for a monopole-antimonopole interaction. For a given monopole configuration with $n^{(a)}$ monopoles and $\bar{n}^{(a)}$ antimonopoles of types $a = 1,\dots,N$, at positions $\mathbf{r}_{k}^{(a)}, k = 1,\dots,n^{a}$ and $\bar{\mathbf{r}}_{k}^{(a)}, k = 1,\dots,\bar{n}^{a}$ respectively, the three dimensional $U(1)^N$ magnetic field is given by
\be \label{complicatedmagneticfield}
	\mathbf{B} (\mathbf{x}) = \sum_{a = 1}^{N} \frac{2 \pi}{g} \alpha_{a} \left[
		\sum_{k = 1}^{n^{(a)}} \frac{\mathbf{x} - \mathbf{r}_{k}^{(a)}}{4 \pi \left| \mathbf{x} - \mathbf{r}_{k}^{(a)} \right|^{3}}
		- \sum_{l = 1}^{\bar{n}^{(a)}} \frac{\mathbf{x} - \bar{\mathbf{r}}_{l}^{(a)}}{4 \pi \left| \mathbf{x} - \bar{\mathbf{r}}_{l}^{(a)} \right|^{3}} \right].
\ee
Letting 
\begin{equation} \label{assumptions}
	\begin{array}{lcl}
		M^{(a)} & = & n^{(a)} + \bar{n}^{(a)} , \\
  		\mathbf{r}_{k}^{(a)} & = & \left\{ \begin{array}{lcl}
			\mathbf{r}_{k}^{(a)} & \mathrm{for} & k \leq n^{(a)} \\
			\bar{\mathbf{r}}_{k-n^{(a)}}^{(a)} & \mathrm{for} & k > n^{(a)} \\
		\end{array} \right. , \\
		Q_{k}^{(a)} & = & \left\{
		\begin{array}{lcl}
			+1 & \mathrm{for} & k \leq n^{(a)} \\
			-1 & \mathrm{for} & k > n^{(a)} \\
		\end{array} \right. , \\
	\end{array}
\end{equation}
we can write (\ref{complicatedmagneticfield}) in a more compact form,
\be \label{magneticfield}
	\mathbf{B} (\mathbf{x}) = \sum_{a = 1}^{N} \frac{2 \pi}{g} \alpha_{a} \left[ \sum_{k = 1}^{M^{(a)}}
		Q_{k}^{(a)} \frac{\mathbf{x} - \mathbf{r}_{k}^{(a)}}{4 \pi \left| \mathbf{x} - \mathbf{r}_{k}^{(a)} \right|^{3} } \right].
\ee
The action for such a monopole configuration is a combination of the monopole self-energies and the Coulomb interaction potential energies for each pair of monopoles,
\be \label{totalaction}
	S_{\mathrm{MG}} = S_{\mathrm{self}} \sum_{a = 1}^{N} M^{(a)} + S_{\mathrm{int}},
\ee
where
\be \label{interaction}
	S_{\mathrm{int}} = \frac{2 \pi^{2} L}{g^{2}} \sum_{a,b = 1}^{N} \alpha_{a} \cdot \alpha_{b} \left[ \sum_{k = 1}^{M^{(a)}}
		\sum_{l=1}^{M^{(b)}} Q_{k}^{(a)} Q_{l}^{(b)} G \left( \mathbf{r}_{k}^{(a)} - \mathbf{r}_{l}^{(b)} \right) \right]
\ee
and
\be \label{greensfunction}
	G(\mathbf{r}) \equiv \frac{1}{4 \pi \left| \mathbf{r} \right|}.
\ee
The canonical partition function is then given, as usual, by a sum over all possible monopole configurations with a statistical weight $e^{-S}$,
\be \label{monopolepartition}
	\mathcal{Z} = \int \prod_{a = 1}^{N} d\mu^{(a)} \; e^{-S_{\mathrm{int}}},
\ee 
with measure
\be \label{monopolemeasure}
	d\mu^{(a)} = \sum_{n^{(a)} = 0}^{\infty} \frac{\left(\zeta/2\right)^{n^{(a)}}}{n^{(a)}!}
		\sum_{\bar{n}^{(a)} = 0}^{\infty} \frac{\left(\zeta/2\right)^{\bar{n}^{(a)}}}{\bar{n}^{(a)}!}
		\int_{\mathbb{R}^{3}} \prod_{k = 1}^{n^{(a)}} d\mathbf{r}_{k}^{(a)}
		\int_{\mathbb{R}^{3}} \prod_{l = 1}^{\bar{n}^{(a)}} d\bar{\mathbf{r}}_{k}^{(a)}.
 \ee
The monopole fugacity, $\zeta$, describes the density of monopoles and is given by,
\be \label{fugacity}
	\zeta \equiv C \; e^{-S_{\mathrm{self}}} = A m_{W}^{3} \left( g^{2} N \right)^{-2} e^{-\Delta S} e^{-8 \pi^{2}/N g^{2}(m_{W})},
\ee
where the $C$ factor is the one-loop functional determinant in the monopole background as described in the appendix of \cite{Yaffe:2008}.

Next we show explicitly that the above monopole partition function (\ref{monopolepartition}) is equivalent to a Sine-gordon partition function which describes the proper $\theta$-dependence for the QCD vacuum.

\subsection{Monopole sine-Gordon equivalence}

The sine-Gordon partition function for this model describes a three dimensional $N$-component real scalar field theory, given by
\be \label{sinegordonpartition}
	\mathcal{Z} = \int \prod_{a = 1}^{N} \mathcal{D}\sigma_{a} \; e^{-S_{\mathrm{dual}}\left[\bm{\sigma}\right]},
\ee
with
\begin{equation} \label{sinegordonaction}
		S_{\mathrm{dual}} = \int_{\mathbb{R}^{3}} d^{3}x \; \left[ \frac{1}{2L}\left(\frac{g}{2\pi}\right)^{2} \left( \nabla \bm{\sigma} \right)^{2}
			- \zeta \sum_{a = 1}^{N} \cos ( \alpha_{a} \cdot \bm{\sigma} ) \right].
\end{equation}
Considering the cosine term,
\begin{equation} \label{cosineterm}
	\begin{array}{lcl}
		\displaystyle \exp \left[ \zeta\int_{\mathbb{R}^{3}} d^{3}x \; \sum_{a = 0}^{N} \cos ( \alpha_{a} \cdot \bm{\sigma} ) \right] & = & 
			\displaystyle \prod_{a = 1}^{N} \exp \left[ \zeta \int_{\mathbb{R}^{3}} d^{3}x \; \cos ( \alpha_{a} \cdot \bm{\sigma} ) \right], \\
		& = & \displaystyle \prod_{a = 1}^{N} \exp \left[ \frac{\zeta}{2} \int_{\mathbb{R}^{3}} d^{3}x \;
			\left( e^{i \alpha_{a} \cdot \bm{\sigma}} + e^{-i \alpha_{a} \cdot \bm{\sigma}} \right) \right], \\
	\end{array}
\end{equation}
we can apply the power series representation for the exponential, $e^{x} = \sum x^{n} / n!$, and get
\be \label{cosineseries}
	(l.h.s.) = \prod_{a = 1}^{N} \left\{ \sum_{M^{(a)} = 0}^{\infty} \frac{(\zeta/2)^{M^{(a)}}}{M^{(a)}!} \prod_{m = 0}^{M^{(a)}}
		\left[ \int_{\mathbb{R}^{3}} d^{3}x_{m} \left( 
		e^{i \alpha_{a} \cdot \bm{\sigma}(x_{m})} + e^{-i \alpha_{a} \cdot \bm{\sigma}(x_{m})} \right )\right] \right\}.
\ee
We then make use of the binomial theorem,
\begin{equation} \label{binomialtheorem}
	(x + y)^{n} = \sum_{k = 0}^{n} \left( 
		\begin{array}{c}
			n \\ k \\
		\end{array}
		\right) x^{n-k}y^{k} \; \mathrm{with} \;
		\left( 
		\begin{array}{c}
			n \\ k \\
		\end{array}
		\right) = \frac{n!}{(n-k)!k!},
\end{equation}
and arrive at,
\begin{equation} \label{simplifiedcosine}
	\begin{array}{lcl}
		(l.h.s.) & = & \displaystyle \prod_{a = 1}^{N} \sum_{M^{(a)} = 0}^{\infty} \left\{
			\sum_{m = 0}^{M^{(a)}} \frac{(\zeta/2)^{M^{(a)}}}{m! (M^{(a)} - m)!} \left[
			\prod_{k = 1}^{M^{(a)}} \int_{\mathbb{R}^{3}} d^{3}x_{k} \right] \right\} \exp \left[
			i \sum_{k = 1}^{M^{(a)}} Q^{(a)}_{k} \alpha_{a} \cdot \bm{\sigma}_{k} \right] \\
		& = &\displaystyle \left[ \prod_{a = 1}^{N} \int d\mu^{(a)} \right] 
			\exp \left[ i \sum_{a = 0}^{N} \sum_{k = 0}^{M^{(a)}} Q_{k}^{(a)} \alpha_{a} \cdot \bm{\sigma} (x_{k}) \right],
	\end{array}
\end{equation}
where $d\mu^{(a)}$ is given in (\ref{monopolemeasure}). Thus, inserting (\ref{simplifiedcosine}) into the sine-Gordon partition function (\ref{sinegordonpartition}), we have
\begin{equation} \label{partwaypartition}
	\mathcal{Z} = \int \prod_{a = 1}^{N} \left[ \mathcal{D}\sigma_{a} \; d\mu^{(a)} \right] \exp \left\{
		- \beta \int_{\mathbb{R}^{3}} d^{3}x \left[ \frac{1}{2} \left( \nabla \bm{\sigma} \right)^{2}
		- \frac{i}{\beta} \sum_{a = 1}^{N} \sum_{k = 1}^{M^{(a)}} Q_{k}^{(a)} 
			\delta (\mathbf{x}_{k}^{(a)} - \mathbf{x}) \alpha_{a} \cdot \bm{\sigma} (\mathbf{x}) \right] \right\},
\end{equation}
where
\begin{equation} \label{beta}
	\beta \equiv \frac{1}{L} \left( \frac{g}{2\pi} \right)^{2}.
\end{equation}
Treating the last term in the exponent as a source term,
\begin{equation} \label{sourcedefinition}
	J(\mathbf{x}) \equiv \frac{- i}{\beta} \sum_{a = 1}^{N} \sum_{k = 1}^{M^{(a)}} Q_{k}^{(a)}
		\delta (\mathbf{x}_{k}^{(a)} - \mathbf{x}) \alpha_{a},
\end{equation}
and completing the square with the shift $\sigma (\mathbf{x}) \rightarrow \sigma (\mathbf{x}) + \int d^{3}y \; G(\mathbf{x}-\mathbf{y}) J(\mathbf{y})$, we have,
\begin{equation} \label{seperatedpartition}
	\mathcal{Z} = \mathcal{Z}_{0} \int \left[ \prod_{a = 0}^{N} d\mu^{(a)} \right] \exp \left\{ \frac{\beta}{2}
		\int_{\mathbb{R}^{3}} d^{3}x \int_{\mathbb{R}^{3}} d^{3}y
		\left[J(\mathbf{x}) G(\mathbf{x}-\mathbf{y}) J(\mathbf{y}) \right] \right\},
\end{equation}
in which $\mathcal{Z}_{0}$ is the functional determinant
\be \label{determinantfactor}
	\mathcal{Z}_{0} \equiv \int \left[ \prod_{a = 0}^{N} \mathcal{D} \sigma_{a} \right] \exp \left[ \frac{-\beta}{2}
		\int_{\mathbb{R}^{3}} d^{3}x \left( \nabla \bm{\sigma} \right)^{2} \right].
\ee
The above determinant does not contain any of the relevant physics and is just a constant prefactor that will drop out of any calculation of operator expectation values in the monopole ensemble. Finally, inserting the expression for the source (\ref{sourcedefinition}), the partition function becomes,
\begin{equation} \label{finalsteppartition}
	\mathcal{Z}  =  \mathcal{Z}_{0} \int \left[ \prod_{a = 0}^{N} d\mu^{(a)} \right] \exp \left[ \frac{-2 \pi^{2} L}{g^{2}}
		\sum_{a,b = 1}^{N} \sum_{k = 1}^{M^{(a)}} \sum_{l = 0}^{M^{(b)}} \alpha_{a} \cdot \alpha_{b} \;
		Q_{k}^{(a)} Q_{l}^{(b)} G ( \mathbf{x}_{k}^{(a)} - \mathbf{x}_{l}^{(b)} ) \right],
\end{equation}
which is the partition function for the monopole gas from (\ref{monopolepartition}).

Next, including a $\theta$-parameter in the Yang-Mills action,
\be \label{thetaincluded}
	S_{\mathrm{YM}} \rightarrow S_{\mathrm{YM}} + i \theta \int_{\mathbb{R}^{3} \times S^{1}} \frac{1}{16 \pi^{2}} \mathrm{tr}
		\left[ F_{\mu\nu} \tilde{F}^{\mu\nu} \right],
\ee
with $\tilde{F}^{\mu\nu} \equiv \epsilon^{\mu\nu\rho\sigma} F_{\rho\sigma}$, multiplies each monopole fugacity by $e^{i \theta / N}$ and antimonopole fugacity by $e^{- i \theta / N}$. In the dual sine-Gordon theory this inclusion is equivalent to shifting the cosine term so that\footnote{We note in passing that there is a typo in \cite{Yaffe:2008} in sine-Gordon representation which is corrected here. Also, it has been stated (wrongly) in ref.\cite{Yaffe:2008}  that the sine-Gordon Lagrangian  is $2\pi$ periodic as a result of a symmetry. This statement  is incorrect, as the claimed symmetry is not in fact a symmetry of the theory, such that $\theta$ parameter enters the Lagrangian as $\theta/N$ as it should. To check this insert $\bm{\sigma} = 0$ and notice that the $\theta$-dependence is explicitly different after the transformation suggested in ~\cite{Yaffe:2008}. }
\be \label{thetaaction}
	S_{\mathrm{dual}} \rightarrow \int_{\mathbb{R}^{3}} \left[ \frac{1}{2 L} \left( \frac{g}{2 \pi} \right)^{2}
		\left( \nabla \bm{\sigma} \right)^{2} - \zeta \sum_{a = 1}^{N} \cos \left( \alpha_{a} \cdot \bm{\sigma}
		+ \frac{\theta}{N} \right) \right]	.
\ee
The $\theta$ parameter enters the effective Lagrangian (\ref{thetaaction}) as $\theta/N$ which is the direct consequence of the fractional topological charges of the monopoles (\ref{topologicalcharge}). Nevertheless, the 
theory is still $2\pi$ periodic. This
  $2\pi$ periodicity of the theory is restored not due to the $2\pi$ periodicity of Lagrangian (\ref{thetaaction}).
  Rather, it is restored as a result of   summation over all branches of the theory when the  levels cross at
   $\theta=\pi (mod ~2\pi)$ and one branch replaces another and becomes the lowest energy state. 
   Indeed, the ground state energy density is determined by minimization of the effective potential (\ref{thetaaction}) 
   when summation $ \sum_{l=0}^{N-1}$ over all branches is assumed in the definition of the canonical partition function (\ref{monopolepartition}). It  is given by
    \be
\label{min}
E_{min} (\theta)=  - \lim_{V \rightarrow 
\infty} \; \frac{1}{VL} \ln \left\{ 
  \sum_{l=0}^{N-1} \exp \left[ 
V\zeta N\cos \left( \frac{ \theta + 2 \pi \, l }{N} \,\right) \right]   \right\},
\ee
where $V$ is 3d volume of the system. Eq. (\ref{min}) shows that in the limit $V\rightarrow\infty$ cusp singularities occur at the values at $\theta=\pi (mod ~2\pi)$    where the lowest energy vacuum state switches from one analytic branch to another one. The first derivative of the vacuum energy, which is proportional to the topological density condensate, is two-valued at these points. This means that whenever $\theta=\pi (mod ~2\pi)$  we stay with two degenerate vacua in the thermodynamic limit. If, on the other hand, the thermodynamic limit is performed for a fixed value of $\theta$, any information on other states is completely lost in eq. (\ref{min}). Correspondingly, the $2\pi$  periodicity in $\theta$ is also lost in infinite volume formulae. 
   We have no chance to know about additional states when we work in the infinite volume limit from the very beginning.
   As a result, usual $V=\infty$ formulae become blind to the very existence of a whole set of different vacua, which is just responsible for restoration of the $2\pi$  periodicity in $\theta$. The model under consideration explicitly supports this pattern in deformed QCD where all computations are under complete theoretical control.

   Such a pattern is known to emerge in many four dimensional supersymmetric models, and also gluodynamics in the limit $N=\infty$. It has been further argued \cite{Halperin:1997bs} that the same  pattern also emerges in four dimensional gluodynamics at any finite $N$.  We follow, in fact, the   technique from \cite{Halperin:1997bs} to arrive at (\ref{min}) in analyzing the $\theta$ periodicity of the theory. The same pattern emerges in holographic description of QCD ~\cite{wittenflux} at $N=\infty$ as well.
   
Finally, considering the expectation value
\begin{equation} \label{monopolecreationexpectation}
	\begin{array}{lcl}
		\displaystyle \< e^{\pm i \alpha_{b} \cdot \bm{\sigma} (\mathbf{y})} \> & = & \displaystyle
			\frac{1}{\mathcal{Z}} \int \left[ \prod_{a = 1}^{N} \mathcal{D}\sigma_{a} \; d\mu^{(a)} \right]
			\exp \left\{ -\beta \int_{\mathbb{R}^{3}} d^{3}x \left[ \frac{1}{2} \left( \nabla \bm{\sigma} \right)^{2} +
			J \cdot \bm{\sigma} \mp \frac{i}{\beta} \delta (x - y) \alpha_{b} \cdot \bm{\sigma} \right] \right\} \\
		& = & \displaystyle \frac{\mathcal{Z}_{0}}{\mathcal{Z}} \int \left[ \prod_{a = 0}^{N} d\mu^{(a)} \right] e^{-S_{\mathrm{MG}}}
			\exp \left[ \pm \frac{4 \pi^{2} L}{g^{2}} \sum_{a = 1}^{N} \sum_{k = 1}^{M^{(a)}} Q_{k}^{(a)}
			\alpha_{a} \cdot \alpha_{b} G(x_{k}^{(a)} - y) \right], 
	\end{array}
\end{equation}
we note that the operator $e^{i \alpha_{a} \cdot \bm{\sigma} (\mathbf{x})}$ is the creation operator for a monopole of type $a$ at $\mathbf{x}$, i.e.
\be
\label{operator}
{\cal{M}}_a (\mathbf{x}) =e^{i \alpha_{a} \cdot \bm{\sigma} (\mathbf{x})}.
\ee
Likewise,  the operator for an antimonopole is ${\bar{\cal{M}}}_a (\mathbf{x})= e^{-i \alpha_{a} \cdot \bm{\sigma} (\mathbf{x})}$.
The expectation values of these operators in fact determine the ground state of the theory. 

\subsection{Mass gap}\label{gap}

The cosine potential in the sine-Gordon action (\ref{sinegordonaction}) gives rise to a mass term for the dual scalar fields. Expanding the potential around the minimum $\bm{\sigma} = 0$ up to quadratic order and rescaling $\bm{\sigma} \rightarrow L (2\pi)^{2} / g^{2} \bm{\sigma}$ to put the kinetic term into canonical form, gives (up to a constant term)
\be \label{expandedpotential}
	V(\bm{\sigma}) \cong \frac{1}{2} m_{\sigma}^{2} \sum_{a = 1}^{N} (\sigma_{a+1} - \sigma_{i})^{2},
\ee
with
\be \label{sigmamass}
	m_{\sigma}^{2} \equiv L \zeta \left( \frac{2\pi}{g} \right)^{2}.
\ee
The above mass term is diagonalized by the discrete Fourier transform
\be \label{discretefourier}
	\tilde{\sigma}_{b} \equiv \frac{1}{\sqrt{N}} \sum_{a = 0}^{N} e^{\frac{-2 \pi i a b}{N}} \sigma_{a},
\ee
becoming
\be \label{diagonalmass}
	V(\bm{\sigma}) \cong \frac{1}{2} \sum_{a = 1}^{N} m_{a}^{2} \left| \tilde{\sigma}_{a} \right|^{2},
\ee
where $m_{a} = m_{\sigma} \sin (\pi a / N)$. So the only scalar field which remains massless is the $N$th field, which is the field associated with the affine root. Inserting the discrete Fourier transform (\ref{discretefourier}) into the full cosine potential however shows that the $N$th field drops out of the cosine potential completely, so although it remains massless, it completely decouples from the theory and does not interact with the other components at all.

\section{Topological susceptibility in the deformed QCD}\label{section-chi}

Next we consider the topological susceptibility in the deformed theory discussed in the previous section in both the monopole and sine-Gordon formalisms. We define the topological density $q(\mathbf{x})$ and topological charge $Q$ by
\begin{equation} \label{topologicalcharge1}
	Q \equiv		\frac{1}{16 \pi^2} \int_{\mathbb{R}^{3} \times S^1} d^4x \; \mathrm{tr} \left[ F_{\mu\nu} \tilde{F}^{\mu\nu} \right] =L \int_{\mathbb{R}^{3}} d^{3}x \; \left[ q (\mathbf{x}) \right] 
\end{equation}
and as in (\ref{chi}) the topological susceptibility $\chi$ is given by
\begin{equation} \label{deformedchi}
	\chi = L\lim_{\mathbf{k} \rightarrow 0} \int d^3x \; e^{i \mathbf{k} \cdot \mathbf{x}} \left< q(\mathbf{x}) q(0) \right>.
\end{equation}
First, in next subsection we compute the topological susceptibility directly, using the monopole gas representation. As the next step, we reproduce our results using sine Gordon representation of the theory. Finally, we compute the topological susceptibility when a single massless quark is introduced into the system. Essentially, the goal here is to discuss the same physics related to non-dispersive contact term, topological sectors and all that in deformed QCD, in close analogy to our discussions in 2d QED in section \ref{2d}. 

\subsection{Topological susceptibility in the monopole picture}\label{section-monopoles}

In order to compute the functional form of the topological susceptibility in the monopole theory we consider the topological density,
\begin{equation} \label{topologicaldensitydefinition}
	\begin{array}{lcl}
		q(\mathbf{x}) & = & \displaystyle  \frac{1}{16 \pi^2} \mathrm{tr} \left[ F_{\mu\nu} \tilde{F}^{\mu\nu} \right] 
			 =  \displaystyle \frac{-1}{8 \pi^2} \epsilon^{i j k 4} \sum_{a = 1}^{N} F^{(a)}_{j k} F^{(a)}_{i 4} \\
			& = & \displaystyle \frac{g }{4 \pi^2} \sum_{a = 1}^{N} \left< A_{4}^{(a)} \right>
				\left[ \nabla \cdot \mathbf{B}^{(a)} (\mathbf{x}) \right],
	\end{array}
\end{equation}
where the $U(1)^N$ magnetic field, $B^i = \epsilon^{ijk4} F_{jk}/2g$ is given by
\begin{equation} \label{magneticfieldpart}
	\mathbf{B}^{(a)} (\mathbf{x}) = \frac{2\pi}{g} \alpha_{a} \left[ \sum_{k = 1}^{n^{(a)}}
		\frac{\mathbf{x} - \mathbf{r}_{k}^{(a)}}{4\pi \left| \mathbf{x} - \mathbf{r}_{k}^{(a)} \right|^3}
		-\sum_{k = 1}^{n^{(a)}} \frac{\mathbf{x} - \mathbf{r}_{k}^{(a)}}
		{4\pi \left| \mathbf{x} - \mathbf{r}_{k}^{(a)} \right|^3} \right],
\end{equation}
and $\left< A_4^{(a)} \right>$ is just the expectation value of the diagonal gauge fields in the compact direction,
\begin{equation} \label{higgsexpectation}
	\left< A_4^{(a)} \right> = \frac{2\pi}{N L}\mu^a.
\end{equation}
The above $\mu^a$ are the fundamental weights for the $SU(N)$ algebra and are defined by
\begin{equation} \label{fundamentalweights}
	\mu^a \cdot \alpha_{b} \equiv \frac{1}{2} \delta^{a}_{b} \alpha^{2}_{b} = \delta^{a}_{b}.
\end{equation}
Inserting the magnetic field for the monopole ensemble into the topological density expression (\ref{topologicaldensitydefinition}) and applying Gauss's theorem to the result, we arrive at
\begin{equation} \label{topologicaldensity}
	\begin{array}{lcl}
		q (\mathbf{x}) & = & \displaystyle \sum_{a = 1}^{N} \frac{1}{LN} \left[ \sum_{k = 1}^{n^{(a)}} \delta (\mathbf{r}_{k}^{(a)} - \mathbf{x})
			-\sum_{l = 1}^{\bar{n}^{(a)}} \delta (\bar{\mathbf{r}}_{l}^{(a)} - \mathbf{x}) \right] \\
		 & = & \displaystyle \frac{1}{LN} \sum_{a = 1}^{N} \sum_{k = 1}^{M^{(a)}} Q_{k}^{(a)} \delta (\mathbf{r}_{k}^{(a)} - \mathbf{x}),
	\end{array}
\end{equation}
which  obviously gives the proper topological charge for a single monopole or antimonopole, $Q = \pm1/N$. The topological density operator $q (\mathbf{x})$ has dimension four as it should. 

The expectation value $\left< q(\mathbf{x}) q(0) \right>$ is the topological density operator (\ref{topologicaldensity}) evaluated at each point inserted in the partition function (\ref{monopolepartition}),
\begin{equation} \label{qqexpectation}
	\begin{array}{lcl}
		\left< q(\mathbf{x}) q(0) \right> & = & \displaystyle \frac{1}{{\mathcal{Z}}} \int \prod_{a = 1}^{N} d\mu^{(a)} \;
			\left[ q(\mathbf{x}) q(0) \right] e^{-S_{\mathrm{int}}} \\
		& = & \displaystyle \frac{1}{\mathcal{Z} N^2L^2} \int \prod_{a = 1}^{N} d\mu^{(a)} \sum_{a,b = 1}^{N}
			\sum_{k = 1}^{M^{(a)}} \sum_{l = 1}^{M^{(b)}} \left[ Q_{k}^{(a)} Q_{l}^{(b)}
				\delta (\mathbf{r}_{k}^{(a)} - \mathbf{x}) \delta(\mathbf{r}_{l}^{(b)}) \right] e^{-S_{\mathrm{int}}} \\
		& = & \displaystyle \frac{1}{\mathcal{Z} N^2L^2} \int d\mu \sum_{m} \sum_{n}
			\left[ Q_{m} Q_{n} \; \delta (\mathbf{r}_{m} - \mathbf{x}) \delta(\mathbf{r}_{n}) \right] e^{-S_{\mathrm{int}}} \\
		& = & \displaystyle \frac{1}{\mathcal{Z} N^2L^2} \int d\mu \left\{ \delta(\mathbf{x}) \sum_{m} \delta (\mathbf{r}_{m}) e^{-S_{\mathrm{int}}}
			+ \sum_{m} \sum_{n \neq m} \left[ Q_{m} Q_{n} \;
			\delta (\mathbf{r}_{m} - \mathbf{x}) \delta(\mathbf{r}_{n}) \right] e^{-S_{\mathrm{int}}} \right\} \\
		& = & \displaystyle \frac{\zeta}{NL^2} \left\{ \delta (\mathbf{x})
			- \cal{O}( {\zeta})  \right\},
	\end{array}
\end{equation}
where we have condensed the indices to just $m$ and $n$ which run over each monopole in the ensemble. In the above expression, the double sum of delta functions gives a set of terms in which each pair of monopoles in the ensemble are moved to the points $\mathbf{x}$ and $0$ and computes the partition function given that arrangement. The monopole gas experiences Debye screening so that the field due to any static charge falls off exponentially with characteristic length $m_{\sigma}^{-1}$. The number density $\cal{N}$ of monopoles is given by the monopole fugacity, $\sim \zeta$, so that the average number of monopoles in a ``Debye volume" is given by
\begin{equation} \label{debye}
{\cal{N}}\equiv	m_{\sigma}^{-3} \zeta = \left( \frac{g}{2\pi} \right)^{3} \frac{1}{\sqrt{L^3 \zeta}} \gg 1.
\end{equation} 
The last inequality holds since the monopole fugacity is exponentially suppressed, $\zeta \sim e^{-1/g^2}$, and in fact we can view (\ref{debye}) as a constraint on the validity of our approximation. The statement here is that inserting or removing a particular monopole will not drastically affect the monopole ensemble as a result of condition (\ref{debye}), so that we can compute expectations of operators in the original ensemble without considering the back-reaction on the ensemble itself. 
\exclude{Then, because removing any given monopole does not significantly change the ensemble, we can treat the delta functions in the third line of (\ref{qqexpectation}) as simply creation operators. The second term in (\ref{qqexpectation}), which is a dispersive term, reduces to the form $\< M^{\dag} M\>$ since it is only non-zero for monopole-antimonopole pairs of the same type.}
The factor overall factor of $\zeta$ and additional factor in $ \cal{O}( {\zeta})$ in formula (\ref{qqexpectation}) appears because each monopole we remove from the ensemble leaves a factor of $\zeta /N^{(a)}$ in the monopole measure, and there are $N^{(a)}$ such terms for each type of monopole so that we are left with just a factor of $\zeta$.

The computed non-dispersive contribution (\ref{qqexpectation}) to the topological susceptibility in deformed QCD  has exactly the same structure we observed  in two dimensional  QED discussed in section \ref {2d}. In particular, it is expressed in terms of a $\delta (\mathbf{x})$ function, and it has the ``wrong sign" similar to (\ref{exact1}). Furthermore, this contribution  is not related to any physical propagating degrees of freedom, but rather, it is determined by degenerate topological sectors of the theory. The corresponding ``degeneracy" is formulated in terms of monopoles which essentially describe the tunnelling transitions between those ``degenerate" sectors, see section \ref{interpretation} for more comments on the physical meaning of the formula. If we neglect a small term  $ \cal{O}( {\zeta})$ in formula (\ref{qqexpectation}) we arrive to the following final expression for the topological susceptibility in deformed QCD without quarks 
\be
\label{chi-deformed}
\chi_{YM} = \frac{\zeta}{NL^2} \int d^4 x \left[ \delta(\mathbf{x})
		  \right]=  \frac{\zeta}{NL}.
\ee
It has dimension four as it should. This expression  is a direct analog of eq. (\ref{exact1}) derived for two dimensional  QED.
The same formula (\ref{chi-deformed}) could be computed much easily  in the dual sine-Gordon theory by differentiation of the ground state energy density (\ref{min}) with respect to the $\theta$ as general expression (\ref{chi}) states
\be	\label{chi-YM}
  \chi_{YM} (\theta=0) =  \left. \frac{\partial^2E_{\mathrm{min}}(\theta)}{\partial \theta^2} \right|_{\theta=0}=  \frac{\zeta}{NL}, ~~~~ 
    E_{\mathrm{min}} (\theta=0)= -\frac{N\zeta}{L}.
\ee
Agreement between the two computations  can be considered as a consistency check of our approach in weakly  coupled regime. One can explicitly see that the general relation $\chi_{YM} \sim E_{\mathrm{min}}(\theta=0)/N^2$ holds for the deformed QCD as a result of $\theta/N$ dependence in eq. (\ref{thetaaction}). The difference with real strongly coupled QCD is that the vacuum energy scales as $N^2$ in QCD rather than $\sim N$ in (\ref{chi-YM}).

\subsection{Topological susceptibility in the presence of the light quarks}\label{quarks}

Our goal here is to introduce a single massless quark $\psi$ into the system to see how the topological susceptibility changes in this case. We anticipate that the emerging structure should be very similar to (\ref{exact}) as the topological susceptibility must vanish in the presence of massless quark in  the system:  $\chi_{QCD} (m_q=0)=0$ as the direct consequence of the Ward Identities discussed in section \ref{2d}. 

The low energy description of the system in confined phase with a single quark is accomplished by introducing the $\eta'$ meson. As usual, the $\eta'$ would be conventional massless Goldstone boson if the chiral anomaly is ignored. In the dual sine-Gordon theory   the $\eta'$ field   appears  exclusively in combination with the $\theta$ parameter as $\theta \rightarrow \theta - \eta'$. As it is well known, this is the direct result of the transformation properties of the path integral measure under the chiral transformations $\psi\rightarrow \exp(i \gamma_5\frac{\eta'}{2})\psi$.  Therefore we have,
\be
 \label{etapartition}
	\mathcal{Z} &=& \int \prod_{a = 0}^{N} \mathcal{D} \sigma_{a} \mathcal{D}\eta' \exp\{ -S_{\sigma}-S_{\eta'}-S_{int}\}\\
	S_{\sigma}&=&   \int_{\mathbb{R}^3} d^3x\cdot
		\frac{1}{2 L} \left(\frac{g}{2\pi}\right)^2 \left( \nabla \bm{\sigma} \right)^2  \nonumber\\
		S_{\eta'} &=& 
		  \int_{\mathbb{R}^3} d^3x\cdot  \frac{c}{2} \left( \nabla \eta' \right)^2  \nonumber\\
			S_{int}&=& -  \int_{\mathbb{R}^3} d^3x \cdot   \zeta \sum_{a = 1}^{N}
			\cos \left( \alpha_a \cdot \bm{\sigma} + \frac{\theta - \eta'}{N} \right),  \nonumber
\ee
where coefficient $c$ determines the normalization of the $\eta'$ field and has dimension one. This coefficient,  in principle, can be computed in this model, but such a computation is beyond the scope of the present work. In four dimensional QCD the coefficient $c$ is expressed in terms of standard notations as $({c}/{L})\rightarrow f_{\eta'}^2$. In terms of these parameters the $\eta'$ mass is given by
\be	\label{eta'}
  m_{\eta'}^2=\frac{\zeta}{c N}.
\ee

Since $\eta'$ shows up in the Yang-Mills Lagrangian as $\eta' F \tilde{F}$, we can compute our requisite expectation, $\< F\tilde{F}, F\tilde{F} \>$, by functional differentiation with respect to $\eta'$,
\begin{equation} \label{differentiation}
	\< q(\mathbf{x}) q(\mathbf{y}) \> = \left. \frac{1}{\mathcal{Z}} \frac{i \; \delta}{\delta \eta'(x)}
		\frac{i \; \delta}{\delta \eta'(y)} \mathcal{Z} \right|_{\theta = 0} .
\end{equation}
Thus we have,
\be
 \label{topologicalexpectation}
	\< q(\mathbf{x}) q(\mathbf{y}) \> & = & \displaystyle \frac{1}{\mathcal{Z}} \frac{\delta}{\delta \eta'(y)}
			\int \mathcal{D}\bm{\sigma}\mathcal{D}\eta' \left[ \frac{-\zeta}{NL^2} \int_{\mathbb{R}^3}
			d^3 r_1 \; \delta(\mathbf{r}_1 - \mathbf{x}) \sum_{a = 1}^{N} \frac{\eta'(\mathbf{r}_1)}{N} \right] e^{-S}\nonumber \\
		& = & \displaystyle \frac{1}{\mathcal{Z}} \int \mathcal{D}\bm{\sigma}\mathcal{D}\eta' \left[
			\frac{\zeta}{NL^2} \int d^3 r_1 \; \delta(\mathbf{r}_1 - \mathbf{x}) \delta(\mathbf{r}_1 - \mathbf{y}) \right] e^{-S} \nonumber \\
		&-&	 \displaystyle \frac{1}{\mathcal{Z}} \int \mathcal{D}\bm{\sigma} \mathcal{D}\eta'\left[
			\frac{\zeta^2}{NL^2} \int d^3 r_1 \int d^3 r_2 \; \delta(\mathbf{r}_1 - \mathbf{x})
			\delta(\mathbf{r}_2 - \mathbf{y}) \sum_{a,b = 1}^{N} \frac{\eta'(\mathbf{r}_1)}{N}
			\frac{\eta'(\mathbf{r}_2)}{N} \right] e^{-S} \nonumber \\
		& = & \displaystyle \frac{\zeta}{NL^2} \left [ \delta(\mathbf{x}-\mathbf{y})
			- \frac{\zeta}{N} \< \eta'(\mathbf{x}) \eta'(\mathbf{y}) \> \right]. 
\ee
The first term in (\ref{topologicalexpectation}) is precisely  non-dispersive contact term with the ``wrong sign'' that we computed previously in pure gauge theory using two different methods, see (\ref{chi-deformed}) and (\ref{chi-YM}). The second term represents the conventional dispersive contribution of the physical $\eta'$ state
\footnote{This additional interactions due to the  $\eta'$ exchange may in fact be used as a probe to study the relevant topological charges present in the system. It was precisely the idea behind the proposal, see relatively recent papers~\cite{Zhitnitsky:2006sr,Parnachev:2008fy} and earlier references therein, that $\theta/N$ behaviour unambiguously implies that the relevant vacuum fluctuations  must have fractional topological charges $1/N$. In the present weakly coupled regime these ideas have a precise realization as the basic vacuum fluctuations  are indeed the fractionally charged monopoles (\ref{topologicaldensity}). The results of ~\cite{Zhitnitsky:2006sr,Parnachev:2008fy} are in fact much more generic as they are not based on a weakness of the interaction or semiclassical expansion, but rather, on generic features of the $\eta'$ system which are unambiguously fixed by WI. In the approach advocated in ~\cite{Zhitnitsky:2006sr,Parnachev:2008fy}  one can not study the dynamics of fractionally charged constituents in contrast with the present paper where the dynamics is completely fixed and governed by (\ref{etapartition}). However, the fact that the constituents carry fractional topological charge $1/N$ can be recovered in the approach ~\cite{Zhitnitsky:2006sr,Parnachev:2008fy} because the color-singlet $\eta'$ field enters the effective Lagrangian in a unique way and serves as a perfect probe of the relevant topological charges of the constituents in the system.}.
 One can compute it by redefining $\eta'\rightarrow \eta'/\sqrt{c}$ field to bring its kinetic term $S_{\eta'}$  to the canonical form. In the lowest order approximation it is reduced to the conventional Green's function of the free massive $\eta'$ scalar field  with mass determined by (\ref{eta'}), such that
\begin{equation} \label{QCD}
	\chi_{QCD} = \int d^4 x \< q(\mathbf{x}) q(\mathbf{y}) \>  =\frac{\zeta}{NL} \int d^3 x \left[ \delta(\mathbf{x})
		-m_{\eta'}^2 \frac{e^{-m_{\eta'}r}}{4\pi r}  \right]=0,
\end{equation}
where we represented the canonical  $\eta'$ propagator in terms of its free Green's function in three dimensions. 

The structure of this equation follows precisely the same pattern we observed in analysis of two dimensional  QED, see (\ref{exact}). Indeed, it contains the non-dispersive term due to the degeneracy of the topological sectors of the theory. This contact term (which is not related to any physical propagating degrees of freedom) has been computed  using  monopoles describing the transitions between these topological sectors (\ref{chi-deformed}). The second term emerges as a result of insertion of  the massless quark into the system. It enters $\chi_{QCD}$ precisely in  such a way that the Ward Identity $\chi_{QCD} (m_q=0)=0$ is automatically satisfied as a result of cancellation between the two terms in close analogy with  two dimensional case (\ref{chi2}). We should also mention that very similar structure emerges in real strongly coupled QCD in the framework where the contact term is saturated by the Veneziano ghost. This structure has been confirmed by the QCD lattice studies, see \cite{Zhitnitsky:2010zx,Zhitnitsky:2012im} for the details and references on original lattice results. 

\subsection{Interpretation}\label{interpretation}
 The  results derived in previous sections were formulated in Euclidean space using conventional Euclidean path integral approach.
 Our goal here is to give a physical interpretation of these results in physical  terms  formulated in Minkowski space time.
First of all, the $\delta (\mathbf{x})$ function which appear in the expression for topological susceptibility (\ref{chi-deformed}) should, in fact, be understood as total divergence, 
\be	\label{divergence}
  \chi\sim   \int \delta (\mathbf{x})  \dd^3x = \int   \dd^3x~
  \partial_{\mu}\left(\frac{x^{\mu}}{4\pi x^3}\right)=
    \oint_{S_2}    \dd\Sigma_{\mu}
 \left(\frac{x^{\mu}}{4\pi x^3}\right).
\ee
Indeed, the starting point to derive $\chi$ was the topological density operator (\ref{topologicaldensity}) which is expressed in terms of $\delta (\mathbf{x}-\bold{x_i})$ functions, but in fact represents the   topologically nontrivial boundary conditions determined by the behaviour at a distant surface $S_2$ as (\ref{divergence}) states. The representation (\ref{divergence}) explicitly shows that we are not dealing with ultraviolet (UV) properties of the problem wherein our approximation breaks down. Our treatment of the problem is perfectly justified as $\delta (\mathbf{x}-\bold{x_i})$ functions actually represent the far  infrared (IR) part of physics rather than UV physics. This explains why our description  is valid for $x \gg L$ in spite of presence of the apparently UV singular elements such as the $\delta(\mathbf{x}-\bold{x_i})$ functions which appear in equations of sections \ref{section-monopoles}, \ref{quarks}.

Our next comment is about the interpretation of the classical monopole gas from sections \ref{deformedqcd}. The monopoles in our framework are not real particles, they are pseudo-particles which live in Euclidean space and describe the physical tunnelling processes between different winding states $|n\ra$ and $| n+1 \ra$. The grand canonical partition function written in terms of the classical Coulomb gas (\ref{finalsteppartition}) is simply a convenient way to describe this physics of tunnelling. In particular, the monopole fugacity $\zeta$  together with factor $ L^{-1}$ should be understood as number  of  tunnelling events per unit time per unit volume
\be
\label{zeta}
  \left(\frac{``\cal{N}" {\rm ~of  ~tunnelling ~ events}}{VL}\right)=\frac{N\zeta}{L},
\ee
where extra factor $N$ in (\ref{zeta})  accounts for $N$ different types of monopoles present in the system. The expression (\ref{zeta}) is precisely the contact term, up to factor $1/N^2$ computed in (\ref{chi-deformed}). It is not a coincidence that number of tunnelling events per unit time per unit volume  precisely  concurs  with the absolute value of the energy density of the system (\ref{chi-YM}), since the energy density  (\ref{chi-YM}) in our model is saturated by the topological fluctuations which are not related to any physical propagating degrees of freedom.  We emphasize that while this energy density is not related to any fluctuations of real physical particles, this energy, nevertheless, is still real physical observable parameter, though it can not be defined in terms of conventional Dayson T -product. Instead,  it is defined in terms of the Wick's T product, see Appendix of ref.\cite{Zhitnitsky:2011tr} on a number of subtleties with definition of the energy. 

Finally, the characteristic Debye screening length which appears in the Coulomb gas representation in section \ref{deformedqcd}  
 \be
\label{Debye}
  r_D\equiv m_{\sigma}^{-1}=\frac{g}{2\pi \sqrt{L \zeta}}\gg {L}
\ee
should be interpreted as a typical distance in physical $3d$ space in which the tunnelling event is felt by other fields present in the system. 
The tunnelling  interpretation  also explains the ``wrong" sign in residues of the correlation function  (\ref{chi-deformed}) as we describe the tunnelling in terms of the Euclidean   objects interpolating  between physically equivalent topological sectors  $| n\ra$ rather than the tunnelling of conventional physical degree of freedom between distinct vacuum states  in condensed matter physics.

 \section{Conclusion and future directions}\label{conclusion} 

The main results of this work can be formulated as follows. We studied a number of different ingredients related to $\theta$ dependence, the non-dispersive contribution in topological susceptibility with the ``wrong sign'', topological sectors in gauge theories, and related subjects using a simple ``deformed QCD". This model is a weakly coupled gauge theory, which however has all the relevant essential elements allowing us to study difficult and nontrivial questions which are known to be present in real strongly coupled QCD. Essentially we tested the ideas related to the $U(1)_A$ problem formulated long ago in  \cite{witten,ven,vendiv,Rosenzweig:1979ay,Nath:1979ik,Kawarabayashi:1980dp} in a theoretically controllable manner using the deformed QCD as a toy model. One can explicitly see microscopically how all the crucial elements work.

As this model is a weakly coupled gauge theory, one can try to formulate (and answer) many other questions which are normally the prerogative of numerical Monte Carlo simulations. One such question is the study of scaling properties of the contact term. We can address what happens to the contact term when the Minkowski space-time $R_{3,1}$ gets slightly deformed. For example, what happens when infinite Minkowski space-time $R_{3,1}$ is replaced by a large, but finite size torus? Or, what happens when the Minkowski space-time $R_{3,1}$ is replaced by FRW metric characterized by the dimensional parameter $R\sim H^{-1}$ describing the size of horizon ($H$ here is the Hubble constant)? A naive expectation based on common sense suggests that any physical observable in QCD must not be sensitive to very large distances $\sim \exp(-\Lqcd R)$ as QCD has a mass gap $\sim \Lqcd$. Such a naive expectation formally follows from the dispersion relations similar to (\ref{G}) which dictate that a sensitivity to very large distances must be exponentially suppressed when the mass gap is present in the system, and there are no any physical massless states in the spectrum. However, as we discussed in this paper, along with conventional dispersive contribution (\ref{G}) in the system, there is also the non-dispersive contribution (\ref{top1}) which emerges as a result of topologically nontrivial sectors in four dimensional QCD. This contact term may lead to a power like corrections $ R^{-1} +{\cal O} (R)^{-2}$ rather than exponential like $\exp(-\Lqcd R)$ because the dispersion relations do not dictate the scaling properties of this term. In fact, this term in ``deformed QCD"   in infinite Minkowski space has been explicitly computed in this paper and it is given by (\ref{chi-deformed}), (\ref{QCD}). As our model is a weakly coupled gauge theory, one can, in principle, compute the correction to the formulae (\ref{chi-deformed}), (\ref{QCD}) due to the finite size of the system \cite{TZ}. In different words, one can try to compute the corrections to the monopole fugacity $\zeta$ when the model is formulated in a finite manifold determined by size $R$.  We suspect that the correction may demonstrate a Casimir power like behaviour  $R^{-1} +{\cal O} (R)^{-2}$ rather than exponential like behaviour $\exp(-\Lqcd R)$ similar to the previously studied case of two dimensional QED ~\cite{Urban:2009wb} wherein all relevant ingredients are present in the system as discussed in section \ref{2d}. Such a Casimir type behaviour may have profound consequences for cosmology as discussed in \cite{Zhitnitsky:2011tr}.

\section*{Acknowledgements}
  This research was supported in part by the Natural Sciences and Engineering Research Council of Canada.
         

\end{document}